\documentclass{JHEP3}

\usepackage{epsfig,multicol}

\newcommand\fverb{\setbox\pippobox=\hbox\bgroup\verb}
\newcommand\fverbdo{\egroup\medskip\noindent%
			\fbox{\unhbox\pippobox}\ }
\newcommand\fverbit{\egroup\item[\fbox{\unhbox\pippobox}]}
\newbox\pippobox

\newcommand{\tr}{{\textrm {tr}}}
\newcommand{\Tr}{{\textrm {Tr}}}
\newcommand{\Det}{{\textrm {Det}}}

\newcommand{\thru}[1]{\mathrel{\mathop{#1\!\!\!/}}}
\newcommand{\thruu}[1]{\mathrel{\mathop{#1\!\!\!\!/}}}

\title{Polyakov loop at low and high temperatures}

\author{E. Meg\'{\i}as, E. Ruiz Arriola and L.L. Salcedo\\
	Departamento de F\'{\i}sica Moderna, Universidad de Granada, E-18071 
        Granada, Spain\\
	E-mails: \email{emegias@ugr.es,earriola@ugr.es,salcedo@ugr.es}}

\preprint{}

\abstract{We describe how the coupling of the gluonic Polyakov loop to
quarks solves different inconsistencies in the standard treatment of
chiral quark models at finite temperature at the one quark loop
level. Large gauge invariance is incorporated and an
effective theory of quarks and Polyakov loops as basic degrees of
freedom is generated. From this analysis we find a strong suppression of finite
temperature effects in hadronic observables below the deconfinement
phase transition triggered by approximate triality conservation in a
phase where chiral symmetry is spontaneously broken (Polyakov
cooling). We also propose a simple phenomenological model to describe
the available lattice data for the renormalized Polyakov loop in the
deconfinement phase. Our analysis shows that non perturbative
contributions driven by dimension-2 gluon condensates dominate the
behaviour of the Polyakov loop in the regime $T_c < T < 6T_c$.  }


\dedicated{Talk given at the 29th Johns Hopkins Workshop in Theoretical Physics,\\
 Budapest, August 1-3 2005.}

\begin{document} 


\section{Introduction}
\label{sec:introduction}

Pure gluodynamics formulated using the imaginary time formalism of
finite temperature field theory has an extra discrete glogal symmetry
${\mathbb Z}(N_c)$, which is the center of the usual gauge group
SU($N_c$). A natural order parameter for the transition from the
confining phase, where ${\mathbb Z}(N_c)$ is preserved, to the
deconfining phase, where this symmetry is spontaneously broken, is the
traced Polyakov loop (for a comprehensive review see
e.g.~\cite{Pisarski:2002ji}), defined by
\begin{equation}
L(T)= \langle \tr \; \Omega(x) \rangle =\Big\langle \frac{1}{N_c} \tr \;{\mathbf P} \left( e^{ig\int_0^{1/T} dx_0 A_0(\mathbf{x},x_0)} \right)\Big\rangle \,,
\label{eq:def_polyakov_loop}
\end{equation}
where $\langle \; \rangle$ denotes vacuum expectation value. $A_0$ is
the gluon field in the (Euclidean) time direction. There have been many
efforts in studying the Polyakov loop. A perturbative evaluation of the
Polyakov loop was carried out long ago~\cite{Gava:1981qd} at high
temperatures. An update of this calculations was presented by us in
\cite{Megias:2003ui}. Different renormalization procedures have been
proposed on the lattice simulations more recently
\cite{Kaczmarek:2002mc,Dumitru:2003hp}.

In full QCD, i.e. with dynamical fermions, the Polyakov loop appears
to be an approximate order parameter, as lattice simulations suggest
\cite{Kaczmarek:2005ui}. This may look a bit puzzling since the center
symmetry is largely broken for current quarks. However, as discussed
in \cite{Fukushima:2003fw} in the context of chiral quark models, the
relevant scale stemming from the fermion determinant is in fact the
constituent quark mass, generated by spontaneous chiral symmetry
breaking. Thus, one expects large violations of the center symmetry to
correlate with chiral symmetry restoration. 

We have analyzed the role of large gauge symmetry in a similar
framework~\cite{Megias:2004hj} yielding a unique way of coupling the
polyakov loop to effective constituent quarks. In
Ref.~\cite{Megias:2005ve}, we have also proposed a model to describe
the available lattice data for the renormalized Polyakov loop in terms
of the dimension-2 gluon
condensate~\cite{Chetyrkin:1998yr,RuizArriola:2004en}. This model also
describes consistently the lattice results for the free
energy~\cite{Megias:2005pe}.

\section{Large gauge transformations}
\label{sec:lgt}

In the Matsubara formalism of quantum field theory at finite
temperature the space-time manifold becomes a topological cylinder. In
principle, only periodic gauge transformacions are acceptable since
the quark and gluon fields are stable under these transformations:
\begin{equation}
g(\vec{x},x_0) = g(\vec{x},x_0+\beta) \,,
\label{eq:gauge_tr_per}
\end{equation}
where $\beta=1/T$. In the Polyakov gauge, where $\partial_0 A_0 =0$, $A_0$ is a diagonal and traceless $N_c \times N_c$  matrix. We can consider the following periodic gauge transformation
\begin{equation}
g(x_0)=e^{i2\pi x_0 \Lambda / \beta} \,,
\label{eq:gauge_tr_pl}
\end{equation}
where $\Lambda$ is a color traceless diagonal matrix of integers. Note
that it cannot be considered to be close to the identity, and in that
sense we call it a large gauge transformation. The gauge
transformation on the $A_0$ component of the gluon field is
\begin{equation}
A_0 \rightarrow A_0 + \frac{2\pi}{\beta} \Lambda \,,
\end{equation}
and so gauge invariance manifests as the periodicity of the diagonal
amplitudes of $A_0$ of period $2\pi/\beta$. This invariance is
manifesly broken in perturbation theory, since a periodic function is
approximated by a polynomial.  Nevertheless, we can implement this
large gauge symmetry by considering the Polyakov loop or untraced
Wilson line as an independent degree of freedom, $\Omega(x)$, which
transforms covariantly at $x$
\begin{equation}
\Omega(x) \rightarrow g^{-1}(x) \Omega(x) g(x)\,,
\end{equation}
and, in the Polyakov gauge, $ \Omega(x) = e^{i \beta A_0(\vec{x})} $,
it becomes gauge invariant.

Fermions break the center symmetry of the gauge group, which is
present in all the pure gauge theories. That means that we can only
consider periodic gauge transformations (see
Eq.~({\ref{eq:gauge_tr_per}})). In pure gluodynamics at finite
temperature one can smooth this condition, and consider aperiodic
gauge transformations:
\begin{equation}
g(\vec{x},x_0+\beta) = z g(\vec{x},x_0)  \,, \qquad z^{N_c} = 1 \,.
\end{equation}
Note that $z$ is not an arbitrary phase but an element of ${\mathbb Z}(N_c)$. An example of such a transformation in the Polyakov gauge es given by
\begin{equation}
g(x_0) = e^{i2\pi x_0 \Lambda/N_c\beta} \,,
\end{equation} 
for which $z=e^{i2\pi/N_c}$. The corresponding gauge transformation of the $A_0$ field and the Polyakov loop $\Omega$ is
\begin{equation}
A_0 \rightarrow A_0 + \frac{2\pi}{N_c \beta} \Lambda \,, \qquad
\Omega \rightarrow z \Omega
\label{eq:cst}
\end{equation}
We observe that $\Omega$ transforms as the fundamental representation of the ${\mathbb Z}(N_c)$ group. From Eq.~(\ref{eq:cst}) we deduce that $\langle \Omega \rangle = z \langle \Omega \rangle $ and hence $\langle \Omega \rangle = 0$ in the center symmetric or confining phase. More generally, in this phase
\begin{equation}
\langle \Omega^n \rangle = 0 \qquad {\rm for} \qquad n \ne m N_c \,,
\label{eq:tri_cons}
\end{equation}
with $m$ and arbitrary integer. Obviously, in full QCD the fermion
determinant changes the selection rule Eq.~(\ref{eq:tri_cons}). This
violation is large for massless quarks since we expect corrections $
\sim e^{-M_0/T} \tr \Omega $, and the usefulness of the center
symmetry becomes doubtful.

\section{Problems with Chiral Quark Models at finite temperature}
\label{sec:problems}

The standard treatment of Chiral quark models at finite temperature
presents some inconsistencies. To illustrate this point we will use
the Nambu--Jona-Lasinio model. In the Matsubara formalism we have the
standard rule to pass from $T=0$ formulas to $T \ne 0$, 
\begin{equation}
\int \frac{dk_0}{2\pi} F(\vec{k},k_0) \rightarrow iT \sum_{n=-\infty}^{\infty} F(\vec{k},i \omega_n) \,,
\label{eq:ruleT}
\end{equation}
where $\omega_n=2\pi T(n+1/2)$ are the fermionic Matsubara
frecuencies. The chiral condensate at finite
temperature at one loop level with that rule is given by 
\begin{equation}
\langle \overline{q}q \rangle = 4MT \Tr_c \sum_{\omega_n} 
\int \frac{d^3k}{(2\pi)^3} \frac{1}{\omega_n^2+k^2+M^2} \,,
\end{equation}
where $M$ is the constituent quark mass. Doing the integral and after
a Poisson resummation, we have the low temperature behaviour
\begin{eqnarray}
\langle \overline{q}q \rangle &=& \langle \overline{q}q \rangle_{T=0} -2\frac{N_cM^2 T}{\pi^2} \sum_{n=1}^\infty \frac{(-1)^n}{n} K_1(nM/T)  \nonumber \\
&\stackrel{\rm Low \; T} \sim &\langle \overline{q} q \rangle_{T=0} - \frac{N_c}{2} \sum_{n=1}^\infty  (-1)^n \left( \frac{2MT}{n\pi} \right)^{3/2} e^{-nM/T} \,,
\label{eq:cond_pois}
\end{eqnarray}
where we have used the asymptotic form of the Bessel function
$K_n(z)$. This formula can be interpreted in terms of the quark
propagator in coordinate space
\begin{equation}
S(x) = \int \frac{d^4k}{(2\pi)^4} \frac{e^{-i k \cdot x}}{\thru{k} - M} =  
(i\thru\partial + M) \frac{M^2}{4\pi^2 i } \frac{K_1(\sqrt{-M^2 x^2})}{\sqrt{-M^2 x^2}} \,,
\end{equation} 
so that at low temperature we get $S(\vec{x},i\beta) \stackrel{\rm Low
\; T}\sim e^{-M/T}$, which represents the exponential suppression
for a single quark at low $T$. This means that we can write the quark
condensate in terms of Boltzmann factors with mass $M_n=n M$. We have
the general result that to any observables which are color singlets,
quark models calculations at finite temperature in the one loop
aproximation generate all possible quark states, i.e.
\begin{equation}
{\cal O}^T = {\cal O}^{T=0} + {\cal O}_q e^{-M/T} + {\cal O}_{qq} e^{-2M/T} + \dots 
\end{equation}
Note that while the term ${\cal O}_q$ corresponds to a single quark
state, the next term ${\cal O}_{qq}$ must be a $qq$ diquark state,
corresponding to a single quark line looping twice around the thermal
cylinder. It cannot be a $\overline{q}q$ meson state because at one
loop this state comes from the quark like going upwards and then
downwards in imaginary time, so that the path does not wind around the
thermal cylinder and then it is already counted in the zero
temperature term ${\cal O}^{T=0}$. From Eq.~(\ref{eq:cond_pois}) we
obtain
\begin{equation}
\langle \overline{q} q \rangle_T = \sum_{n=-\infty}^\infty (-1)^n 
\langle \overline{q}(x_0) q(0)\rangle |_{x_0=in\beta} \,.
\label{eq:qq_sum}
\end{equation}
Under a gauge transformation of the central type we have $\overline{q}(n\beta)q(0) \rightarrow z^{-n} \overline{q}(n\beta)q(0)$. This means that Eq.~(\ref{eq:qq_sum}) is not gauge invariant, and the quark condensate can be decomposed as a sum of irreducible representations of a given triality $n$.

Another problem comes from comparison with chiral perturbation theory at finite temperature. In the chiral limit the leading thermal corrections to the quark condensate for $N_f=2$ are given by
\begin{equation}
\langle \overline{q}q\rangle|_{\rm ChPT} = \langle \overline{q}q\rangle_{T=0} 
\left( 1-\frac{T^2}{8f_\pi^2}-\frac{T^4}{384f_\pi^4}+\cdots \right) \,,
\label{eq:chpt}
\end{equation}
which shows that the finite temperature correction is $N_c$-suppressed
as compared to the zero temperature value. This feature contradicts our
result of Eq.~(\ref{eq:qq_sum}) obtained by using the standard finite
temperature treatment of chiral quark models. The problem is that
chiral quark model phase transitions located at $T_c \sim 240 {\rm
MeV}$ are based on such a formula, where multiquark states are excited
even at very low temperatures.

\section{The Polyakov loop chiral quark model}
\label{sec:PL}

We can formally keep track of large gauge invariance by coupling
gluons to the model in a minimal way. As we said in section
\ref{sec:lgt}, a perturbative treatment of the $A_0$ component of
gluon field manifestly breaks gauge invariance at finite temperature,
and we need to consider the Polyakov loop as an independent degree of
freedom. It appears naturally in any finite temperature calculation in
the presence of minimally coupled vector fields within a derivative or
heat kernel expansion~\cite{Megias:2003ui,Megias:2002vr}. Our approach
is similar to that of \cite{Fukushima:2003fw}, except that there a
global Polyakov loop is suggested in analogy with the chemical
potential. Instead we consider a local Polyakov loop $\Omega(\vec{x})$
coupled to the quarks~\cite{Megias:2004hj}. From those calculations we
deduce the rule of Eq.~(\ref{eq:ruleT}), but with the modified
fermionic Matsubara frequencies
\begin{equation}
\hat\omega_n = 2\pi T (n+1/2+\hat\nu) \,, \qquad 
\hat\nu = (2\pi i)^{-1} \log \Omega \,,
\end{equation}
which are shifted by the logarithm of the Polyakov loop~$\Omega = e^{i 2\pi \hat\nu}$, i.e.~$\hat\nu(\vec{x}) = A_0(\vec{x})/(2\pi T)$. The effect of such a shift over a finite temperature fermionic propagator starting and ending at the same point is
\begin{equation}
\tilde{F}(x;x) \rightarrow \sum_{n=-\infty}^\infty
(-\Omega(\vec{x}))^n \tilde{F}(\vec{x},x_0+in\beta;\vec{x},x_0) \,,
\label{eq:F_wpl}
\end{equation}
instead of the $(-1)^n$ factor obtained from the standard rule
Eq.~(\ref{eq:ruleT}) after using Poisson's summation formula and
Fourier transformation.~\footnote{This formula can be interpreted
saying that in a quark loop at finite temperature, the quarks pick up
a phase $(-1)$ due to Fermi-Dirac statistics, and a non Abelian
Aharonov-Bohm factor $\Omega$ each time the quarks wind once around
the compactified thermal cylinder.} 
To restore gauge invariance we project onto the colour singlet sector
by integration of gluons according to the QCD dynamics. Effectively
this implies an average over the local Polyakov loop with some
normalized weight $\sigma(\Omega;\vec{x})d\Omega$. Here $d\Omega$ is
the Haar measure of SU($N_c$) and $\sigma(\Omega;\vec{x})$ the
probability distribution of $\Omega(\vec{x})$ in the gauge group. For
a general function $f(\Omega)$, meaning a ordinary funcion $f(z)$
evaluated at $z=\Omega$, we have
\begin{eqnarray}
\left\langle \frac{1}{N_c} \tr_c f(\Omega) \right\rangle &=&
\int_{{\rm SU}(N_c)}\!\!\! d\Omega\,\sigma(\Omega)  \frac{1}{N_c}
\sum_{j=1}^{N_c}f(e^{i\phi_j})
= \int_0^{2\pi}\frac{d\phi}{2\pi}\hat\sigma(\phi) f(e^{i\phi}) \,,
\label{eq:int_f}
\end{eqnarray}
where $e^{i\phi_j}$, $j=1,\ldots,N_c$ are the eigenvalues of $\Omega$ and
\begin{eqnarray}
\hat\sigma(\phi) &:=& 
\int_{{\rm SU}(N_c)}\!\!\! d\Omega \, \sigma(\Omega) \frac{1}{N_c}
\sum_{j=1}^{N_c}2\pi\delta(\phi-\phi_j) \,.
\end{eqnarray}

By applying the aforementioned rules, the chiral quark model coupled
to the Polyakov loop corresponds to simply make the replacement
 \begin{eqnarray}
\partial_0 \to \partial_0 - i A_0   
\end{eqnarray} 
in the Dirac operator for the quarks, and to consider the partition
function
\begin{eqnarray}
Z = \int DU D \Omega  \, e^{i \Gamma_G [\Omega]} e^{i \Gamma_Q [ U , \Omega ]} \,, 
\label{eq:Z_pnjl} 
\end{eqnarray} 
where $U$ is the nonlinearly transforming pion field, $DU$ is the Haar
measure of the chiral flavour group $SU(N_f)_R \times SU(N_f)_L $ and
$D \Omega $ the Haar measure of the colour group $SU(N_c)$, $\Gamma_G
$ is the effective gluon action whereas $\Gamma_Q$ stands for the
quark effective action. If the gluonic measure is left out $A_0=0$ and
$\Omega=1$ we recover the original form of the corresponding chiral
quark model, where there exists a one-to-one mapping between the loop
expansion and the large $N_c$ expansion both at zero and finite
temperature. Equivalently one can make a saddle point approximation
and corrections thereof. In the presence of the Polyakov loop such a
correspondence does not hold, and we will proceed by a quark loop
expansion, i.e. a saddle point approximation in the bosonic field $U$,
keeping the integration on the (constant) Polyakov loop $\Omega$.

By appling these rules to the quark condensate, we deduce in the
quenched approximation
\begin{equation}
\langle \overline{q}q \rangle_T = \sum_{n=-\infty}^\infty
\frac{1}{N_c} \langle \tr_c (-\Omega)^n \rangle \langle
\overline{q}(x_0)q(0)\rangle|_{x_0=in\beta} \,.
\label{eq:qq_wpl}
\end{equation}
From Eq.~(\ref{eq:tri_cons}) we observe that in the confining phase triality is preserved, so that after gluon average Eq.~(\ref{eq:F_wpl}) becomes
\begin{eqnarray}
&& \tilde{F}(x; x) \to \sum_{n=-\infty}^\infty 
\langle (-\Omega(\vec x))^{nN_c} \rangle 
\tilde{F}( \vec x,x_0+inN_c\beta;\vec x,x_0) \,.
\label{eq:F_wpl_aver}
\end{eqnarray}
At sufficiently low temperature the distribution of the Polyakov loop
becomes just the Haar measure, and one can easily deduce the following
result
\begin{eqnarray}
\langle\tr_c(-\Omega)^n\rangle_{{\rm SU}(N_c)}  = \left\{\matrix{ 
N_c\,, & n=0 \label{eq:p1}\\
-1  \,, & n=\pm N_c \label{eq:p2} \\ 
  0 \,, & {\rm otherwise} \label{eq:p3}\\
} \right.
\nonumber
\end{eqnarray}
Taking into account this formula in Eq.~(\ref{eq:qq_wpl}), we observe
that the inclusion of the Polyakov loop not only removes the triality
breaking terms, but also the thermal contributions are $N_c$
suppressed as compared to the zero temperature value, as is expected
from ChPT (see Eq.~(\ref{eq:chpt})). The quark condensate at finite
temperature at one loop level and in quenched approximation is
\begin{equation}
\langle \overline{q}q\rangle_T = 
\langle \overline{q}q\rangle_{T=0} + 
\frac{2M^2 T}{\pi^2N_c}K_1(N_cM/T) + \cdots \stackrel{\rm Low\; T}\sim  \langle \overline{q}q\rangle_{T=0} 
+4\left(\frac{MT}{2\pi N_c} \right)^{3/2} e^{-N_cM/T} \,.
\label{eq:qq_wpl_aver2}
\end{equation}
The dots indicate higher gluonic or sea quark effects. Due to the
exponential suppression, the leading thermal corrections at one quark
loop level starts only at temperatures near the deconfinement phase
transition. We have named this effect Polyakov
cooling~\cite{Megias:2004hj}, because it is triggered by a group
averaging of Polyakov loops. This means that in the quenched
approximation we do not expect any important finite temperature effect
on quark observables below the deconfinement transition, and the
biggest change should come from pseudoscalar loops at low
temperatures. This is precisely what one expects from ChPT.

\begin{figure}[ttt]
\begin{center}
\epsfig{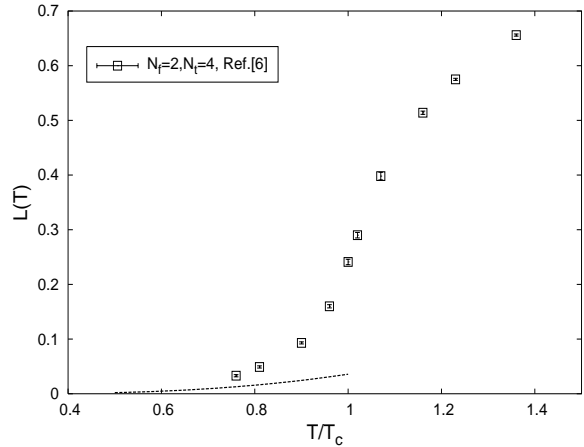}
\end{center}
\caption{Temperature dependence of the renormalized Polyakov loop in
units of the critical temperature. Lattice data correspond to 2-flavor
QCD, and has been taken from~\cite{Kaczmarek:2005ui}. The line
represents our estimation of the Polyakov loop in the low temperature
regime, using $c=3$ as a suitable value for this model-dependent
parameter.}
\label{fig:PLflavor}
\end{figure}

In order to go beyond the quenched approximation, we will consider the
computation of the fermion determinant in the presence of a slowly
varying Polyakov loop following the techniques developped in
\cite{Megias:2002vr}. Such an approximation makes sense in a confining
region where there are very strong correlations beween Polyakov loops.
The fermion determinant can be written as
\begin{equation}
\Det (i\thruu{D}-M) = e^{-\int d^4 y  {\cal L} (y, \Omega) } \,,
\end{equation} 
where ${\cal L}$ is the chiral Lagrangian as a function of the
Polyakov loop which has been computed at finite temperature in
Ref.~\cite{Megias:2004hj} in chiral quark models.  Using this we can
estimate the Polyakov loop~\footnote{The integration can be easily
computed by using the formula $ \langle \tr_c \Omega (x) \; \tr_c
\Omega^{-1} (y) \rangle = e^{-\sigma |x-y| /T} $.  }
\begin{equation}
L = \frac{1}{N_c} \; \frac{ \langle \tr_c \Omega (x) \;
\Det(i\thruu{D}-M) \rangle }{ \langle
\Det(i\thruu{D}-M) \rangle } \stackrel{\rm Low \;T} \sim c
\frac{ 8 \pi T^2 B }{N_c^2 \sigma^3} e^{-M/T} \,,
\label{eq:L_low_T}
\end{equation} 
where $B$ is the vacuum energy density, $\sigma$ is the string tension
and $c$ is a numerical factor which depends on the model. Note that
triality is not preserved due to the presence of dynamical quarks, but
the relevant scale is the constituent quark mass. So the Polyakov loop
can be effectively used as an order parameter. In
Fig.~(\ref{fig:PLflavor}) we confront such an exponential suppression
with unquenched lattice calculations below the phase transition. We
observe that Eq.~(\ref{eq:L_low_T}) could be a good approximation
below $0.6T_c$. In any case, lattice data for lower temperatures are
desirable to do a more precise analysis. For the quark condensate we
take into account the result of Eq.~(\ref{eq:qq_wpl}), so that
\begin{eqnarray}
\langle \overline{q} q \rangle_T = \frac{ \langle \overline{q} q \;
\Det(i\thruu{D}-M) \rangle }{ \langle
\Det(i\thruu{D}-M) \rangle } \stackrel{\rm Low \;T} \sim
\langle\overline{q}q\rangle_{T=0} \times \left( 1+ c' \frac{ 8 \pi T^2 B }{N_c^2 \sigma^3}
e^{-2M/T}\right) \,.
\label{eq:qq_low_T}
\end{eqnarray}
where, again $c'$ depends on the particular model.  In other chiral
quark models, similar results are obtained by replacing $2 M \to M_V $
(the $\rho$ meson mass). The Polyakov cooling persists although is a
bit less effective, and for instance the temperature dependence of the
low energy constants of the effective chiral Lagrangian becomes $
L_i^T - L_i^{T=0} \stackrel{\rm Low \; T} \sim e^{- M_V /T} $.

Finally, on top of this one must include higher quark loops, or
equivalently mesonic excitations. They yield exactly the results of
ChPT~\cite{Florkowski:1996wf} and for massless pions dominate at low
temperatures.  Thus, we see that when suitably coupled to chiral quark
models the Polyakov loop provides a quite natural explanation of
results found long ago on purely hadronic grounds.

\section{Polyakov loop above $T_c$} 
\begin{figure}[tbp]
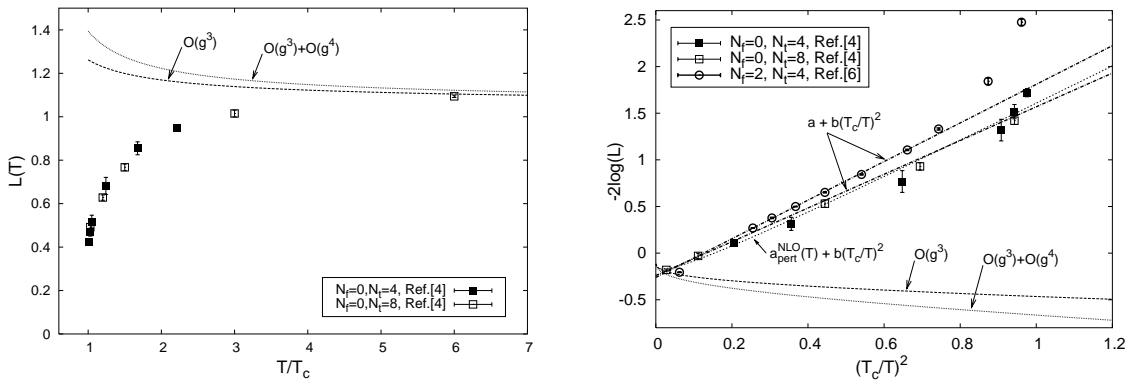

\begin{minipage}[t]{7.3cm}
\includegraphics[width=7.3cm]{figLjhw.epsi}
\end{minipage}
\hspace*{0.3cm}
\begin{minipage}[t]{7.3cm}
\includegraphics[width=7.3cm]{figLLjhw2.epsi}
\end{minipage}
\caption{Temperature dependence of the renormalized Polyakov
 loop. Lattice data from \cite{Kaczmarek:2002mc,Kaczmarek:2005ui}. At
 the left we plot the prediction of perturbation theory at LO and NLO
 in pure gluodynamics~\cite{Gava:1981qd}, and compare with lattice
 data from~\cite{Kaczmarek:2002mc}. At the right, the logarithmic
 dependence of the Polyakov loop versus the inverse temperature
 squared. Fits with $a$ adjustable constant and predicted by NLO
 perturbation theory are displayed. Purely perturbative LO and NLO
 results for $N_f=0$ are shown for comparison.}
\label{fig:log_plot}
\end{figure}

In this section we focus on the behaviour of the Polyakov loop in the
deconfining phase. In that phase chiral symmetry is restored and the
degrees of freedom are quarks and gluons. A perturbative evaluation of
the Polyakov loop was carried out in~\cite{Gava:1981qd} in pure
gluodynamics to NLO, which corresponds to ${\cal O}(g^4)$ in the
Landau gauge. In the Polyakov gauge Eq.~(\ref{eq:def_polyakov_loop})
becomes
\begin{equation}
L(T) = \frac{1}{N_c} \left\langle \tr_c e^{igA_0(\vec{x})/T}\right\rangle =
1-\frac{g^2}{2T^2}\frac{1}{N_c}\langle \tr_c (A_0^2)\rangle 
+ \frac{g^4}{24T^4}\frac{1}{N_c}\langle \tr_c (A_0^4)\rangle + \cdots \,,
\label{eq:pl_pol_gauge}
\end{equation}
where we have considered a series expansion in the gluon
field.~\footnote{Note that the odd order terms vanish due to the
conjugation symmetry of QCD, $A_\mu(x)\rightarrow -A_\mu^T(x)$.} To
describe the dynamics of the $A_0(\mathbf{x})$ field we use the
3-dimensional reduced effective theory of QCD, obtained from the
Euclidean QCD action by integrating the non stationary Matsubara gluon
modes and the quarks~\cite{Megias:2003ui,Appelquist:1981vg}. Let
$D_{00}(\mathbf{k})\delta_{ab}$ denote the 3-dimensional propagator
for the gluon field, then~\footnote{We consider $A_0=\sum_a T_a
A_{0,a}$, with the standard normalization for the Hermitian generators
of the SU($N_c$) Lie algebra in the fundamental representation,
$\tr(T_aT_b)=\delta_{ab}/2$.}
\begin{equation}
\langle A_{0,a}^2\rangle = (N_c^2-1)T \int \frac{d^3k}{(2\pi)^3} D_{00}(\mathbf{k}) \,.
\label{eq:cond}
\end{equation}
To lowest order in perturbation theory the propagator becomes
$D_{00}^P(\vec{k})=1/(\vec{k}^2+m_D^2)$, where $m_D$ is the Debye
mass, which to one loop~\cite{Nadkarni:1982kb} writes $m_D =
gT(N_c/3+N_f/6)^{1/2}$. We can compute $\langle A_0^2 \rangle$ and
$\langle A_0^4 \rangle$ by taking derivatives of the vacuum energy
density of the 3-dimensional theory, already computed to four loops
in~\cite{Kajantie:2003ax}. The contribution from $g^2\langle A_0^4 \rangle$ starts at
${\cal O}(g^6)$, while that of $g^2\langle A_0^2 \rangle$ starts at
${\cal O}(g^3)$. So, the replacement of Eq.~(\ref{eq:pl_pol_gauge})
with
\begin{equation}
L(T) = \exp \left[-\frac{g^2 \langle A_{0,a}^2\rangle}{4N_cT^2} \right] 
\label{eq:gaussian}
\end{equation}
becomes correct up to ${\cal O}(g^5)$. This gaussian ansatz is exact in the large $N_c$ limit. We obtain
\begin{equation}
\langle A_{0,a}^2\rangle^P = -\frac{N_c^2-1}{4\pi} m_D T 
-\frac{N_c(N_c^2-1)}{8\pi^2} g^2 T^2 \left( \log \frac{m_D}{2T}+\frac{3}{4}\right) + {\cal O}(g^3) \,,
\label{eq:A0_pert}
\end{equation}
This result can be deduced in two forms: by derivating the vacuum
energy density~\cite{Kajantie:2003ax}, or by identifying
Eq.~(\ref{eq:gaussian}) with the perturbative result of
Ref.~\cite{Gava:1981qd}. Note that this formula holds also in the
unquenched theory, since to this order, $N_f$ only appears through the
Debye mass. The perturbative contributions to the Polyakov loop at
${\cal O}(g^3)$ and ${\cal O}(g^4)$ have been displayed in
Fig.~(\ref{fig:log_plot}) and compared to the lattice data of the
renormalized Polyakov loop of Ref.~\cite{Kaczmarek:2002mc}. It only
seems to reproduce these data for the highest temperature value
$6T_c$. The results of Ref.~\cite{Kajantie:2003ax} would provide the
${\cal O}(g^5)$ and ${\cal O}(g^6)$ terms. Unfortunately, this
perturbative result is obtained in covariant gauges, generating a
spurious gauge dependence beyond ${\cal O}(g^4)$. In any case these
terms produces a logarithmic dependence with temperature so that their
contribution is similar to that of the lowest orders.

It is clear that perturbation theory cannot explain by itself lattice data, and we propose to account for non perturbative contributions coming from condensates. We consider adding to the propagator new phenomenological pieces driven by positive mass dimension parameters:
\begin{equation}
D_{00}^{NP}(\vec{k}) = \frac{m_G^2}{(\vec{k}^2+m_D^2)^2} \,.
\label{eq:NP_prop}
\end{equation}
This piece produces a non perturbative contribution to the gluon
condensate, namely, $\langle A_{0,a}^2\rangle^{NP} = (N_c^2-1)T m_G^2
/(8\pi m_D)$. If we assume that $m_G$ is temperature independent, the
condensate will also be temperature independent, modulo radiative
corrections. Adding the perturbative and non perturbative
contributions, we get for the Polyakov loop
\begin{equation}
-2\log L = \frac{g^2 \langle A_{0,a}^2\rangle^P}{2N_c T^2} + \frac{g^2 \langle A_{0,a}^2\rangle^{NP}}{2N_c T^2} \,,
\label{eq:log_L}
\end{equation}
where $\langle A_{0,a}^2\rangle^P$ is given by Eq.~(\ref{eq:A0_pert}). We will rewrite this formula as $-2\log L = a + b(T_c/T)^2$.

The lattice data for $-2\log L$ versus $(T_c/T)^2$ are displayed in
Fig.~(\ref{fig:log_plot}). The nearly straight line behaviour is
clear, which means the unequivocal existence of a temperature power
correction driven by a dimension 2 gluon condensate. If we fit the
lattice data by using the perturbative value of $a$ up to NLO,
i.e. ${\cal O}(g^4)$, one obtains:
\begin{equation}
b = \left\{\matrix{
2.27(6)  \,, &&   \cr
2.89(8)  \,, &&   
} \right. 
\qquad
g^2\langle A_{0,a}^2\rangle_T^{NP} = \left\{\matrix{
(1.00(1) \; {\rm GeV})^2 \,, && \qquad\!\!\!\!\!\!\!\!\!\! N_f=0  \cr 
(0.84(2) \; {\rm GeV})^2 \,, && \qquad\!\!\!\!\!\!\!\!\!\! N_f=2
} \right.  \,.
\label{eq:fit_a_NLO}
\end{equation}
In the second case we use only data above $1.1T_c$. A fit of the
lattice data with $a$ treated as a free parameter gives
\begin{equation}
a = \left\{\matrix{
-0.24(1)  \,,    \cr
-0.26(1)  \,, 
} \right. 
\quad
b = \left\{\matrix{
1.81(4)  \,,   \cr
2.07(2)  \,,  
} \right.
\quad
g^2\langle A_{0,a}^2\rangle_T^{NP} = \left\{\matrix{
(0.89(1) \; {\rm GeV})^2 \,, && \quad\!\!\!\!\!\!\!\!\!\! N_f=0  \cr 
(0.71(2) \; {\rm GeV})^2 \,, && \quad\!\!\!\!\!\!\!\!\!\! N_f=2
} \right.  \,.
\label{eq:fit_a_cte}
\end{equation}
Recent analysis of the heavy quark free energy with the model proposed
in Eq.~(\ref{eq:NP_prop}) (see~\cite{Megias:2005pe}) suggest the
possibility that $\alpha_s$ at finite temperature has a smoother
behaviour than the predicted by perturbation theory in the regime
$T_c<T<6T_c$. This is in contrast with existing
analysis~\cite{Kaczmarek:2005ui,Kaczmarek:2004gv}, where authors find
a very large value for $\alpha_s$ in this regime.

We can compare our result for the gluon condensate with finite
temperature determinations based on the study of non perturbative
contributions to the pressure in pure
gluodynamics~\cite{Kajantie:2000iz}. These results yield for the gluon
condensate $(0.93 \pm 0.07\;{\rm GeV})^2$ in the temperature region
used in our fit and in Landau gauge, which is in good agreement with
Eqs.~(\ref{eq:fit_a_NLO}) and (\ref{eq:fit_a_cte}). We also can
compare with zero temperature determinations of the gluon
condensate~$g^2 \langle A_{\mu,a}^2\rangle$ in the Landau gauge and in
quenched QCD. From the gluon propagator~$(2.4 \pm 0.6\;{\rm
GeV})^2$~\cite{Boucaud:2001st} and from the quark propagator~$(2.1 \pm
0.1\;{\rm GeV})^2$~\cite{RuizArriola:2004en}.~\footnote{The total
condensate scales as $D-1$ in the Landau gauge ($D$ is the Euclidean
space dimension).} We observe a remarkable agreement, taking into
account that these results refer to different temperatures and gauges.

\vspace{.5cm}
This work is supported in part by funds provided by the Spanish DGI
and FEDER funds with Grant No. BFM2002-03218, Junta de Andaluc\'{\i}a
Grant No. FQM-225 and EU RTN Contract No. CT-2002-0311(EURIDICE).


\begin{thebibliography}{999}

\bibitem{Pisarski:2002ji}
  R.~D.~Pisarski,
  arXiv:hep-ph/0203271.

\bibitem{Gava:1981qd}
E.~Gava and R.~Jengo,
\newblock Phys. Lett. {\bf B105}, 285 (1981).

\bibitem{Megias:2003ui}
E.~Meg{\'\i}as, E.~Ruiz~Arriola, and L.~L. Salcedo,
\newblock Phys. Rev. {\bf D69}, 116003 (2004).

\bibitem{Kaczmarek:2002mc}
O.~Kaczmarek, F.~Karsch, P.~Petreczky, and F.~Zantow,
\newblock Phys. Lett. {\bf B543}, 41 (2002).

\bibitem{Dumitru:2003hp}
  A.~Dumitru, Y.~Hatta, J.~Lenaghan, K.~Orginos and R.~D.~Pisarski,
  Phys.\ Rev.\ D {\bf 70} (2004) 034511.

\bibitem{Kaczmarek:2005ui}
O.~Kaczmarek and F.~Zantow,
\newblock Phys. Rev. {\bf D71}, 114510 (2005).

\bibitem{Fukushima:2003fw}
  K.~Fukushima,
  Phys.\ Lett.\ B {\bf 591}, 277 (2004).

\bibitem{Megias:2004hj}
E.~Meg\'{\i}as, E.~Ruiz~Arriola and L.~L. Salcedo,
\newblock (2004), arXiv:hep-ph/0412308.

\bibitem{Megias:2005ve}
E.~Meg\'{\i}as, E.~Ruiz~Arriola and L.~L. Salcedo,
\newblock (2005), arXiv:hep-ph/0505215.

\bibitem{Chetyrkin:1998yr}
K.~G. Chetyrkin, S.~Narison, and V.~I. Zakharov,
\newblock Nucl. Phys. {\bf B550}, 353 (1999).

\bibitem{RuizArriola:2004en}
E.~Ruiz~Arriola, P.~O. Bowman, and W.~Broniowski,
\newblock Phys. Rev. {\bf D70}, 097505 (2004).

\bibitem{Megias:2005pe}
  E.~Megias, E.~R.~Arriola and L.~L.~Salcedo,
  arXiv:hep-ph/0510114.

\bibitem{Megias:2002vr}
  E.~Megias, E.~Ruiz Arriola and L.~L.~Salcedo,
  Phys.\ Lett.\ B {\bf 563}, 173 (2003).

\bibitem{Florkowski:1996wf}
  W.~Florkowski and W.~Broniowski,
  Phys.\ Lett.\ B {\bf 386}, 62 (1996).

\bibitem{Appelquist:1981vg}
 T.~Appelquist and R.~D.~Pisarski,
\newblock   Phys. Rev. {\bf D23}, 2305 (1981).

\bibitem{Nadkarni:1982kb}
  S.~Nadkarni,
  Phys.\ Rev.\ D {\bf 27}, 917 (1983).

\bibitem{Kajantie:2003ax}
  K.~Kajantie, M.~Laine, K.~Rummukainen and Y.~Schroder,
  JHEP {\bf 0304}, 036 (2003).

\bibitem{Kaczmarek:2004gv}
O.~Kaczmarek, F.~Karsch, F.~Zantow, and P.~Petreczky,
\newblock  Phys. Rev. {\bf D70}, 074505 (2004).

\bibitem{Kajantie:2000iz}
K.~Kajantie, M.~Laine, K.~Rummukainen, and Y.~Schroder,
\newblock Phys. Rev. Lett. {\bf 86}, 10 (2001).

\bibitem{Boucaud:2001st}
  P.~Boucaud, A.~Le Yaouanc, J.~P.~Leroy, J.~Micheli, O.~Pene and J.~Rodriguez-Quintero,
  Phys.\ Rev.\ D {\bf 63}, 114003 (2001).


\end{thebibliography}
\end{document}